\documentclass[twocolumn,prb,aps,superscriptaddress,showpacs]{revtex4}
\usepackage{graphicx}
\usepackage{epsf}

\begin{document}

\def\bb{\begin{eqnarray}}
\def\ee{\end{eqnarray}}
\title
{Extraction of domain-specific magnetization reversal
for nanofabricated periodic arrays using soft x-ray resonant magnetic scattering}
\author{D. R. Lee}
\email[Electronic mail: ]{drlee@aps.anl.gov}
\author{J. W. Freeland}
\author{G. Srajer}
\affiliation{Advanced Photon Source, Argonne National Laboratory, Argonne, IL 60439}
\author{S. K. Sinha}
\affiliation{Department of Physics, University of California, San Diego, La Jolla, 
CA 92093, \\ and
Los Alamos National Laboratory, Los Alamos, NM 87545}   
\author{V. Metlushko}
\affiliation{Department of Electrical and Computer Engineering, 
University of Illinois at Chicago, Chicago, IL 60607}  
\author{B. Ilic}
\affiliation{School of Applied and Engineering Physics, 
Cornell University, Ithaca, NY 14853}
\date{\today}
\begin{abstract}
A simple scheme to extract
the magnetization reversals of characteristic magnetic domains on
nanofabricated periodic arrays
from soft x-ray resonant magnetic scattering (SXRMS) data
is presented.           
The SXRMS peak intensities from a permalloy square ring array were measured
with field cycling using circularly polarized soft x-rays 
at the Ni L$_3$ absorption edge. 
Various SXRMS hysteresis loops observed at
different diffraction orders enabled the determination of 
the magnetization reversal of each magnetic domain using a simple linear algebra.
The extracted domain-specific hysteresis loops reveal that
the magnetization of the domain parallel to the field
is strongly pinned, while that of the perpendicular domain rotates 
continuously.                
\end{abstract}
\maketitle

Understanding the reversal mechanism of the magnetization in
periodic arrays of submicron and nanoscale magnets is of both
scientific and technological interest.
Fundamental changes in the statics and the dynamics of 
magnetization reversal imposed by nanostructures enrich the 
physics of nanomagnetism.  
A precise control of magnetization reversal involving
well-defined and reproducible magnetic domain states in nanomagnet
arrays is key to future applications, such as high density  
magnetic recording\cite{record_jpd_02} 
or magnetoelectronic\cite{spintronics_ieee_03} devices.
To achieve this, topologically various nanomagnets, ranging from 
simple disks\cite{disk_jpd_00} to more complicated rings\cite{ring_prl_01}
or negative dots (holes)\cite{hole_apl_97},
have been investigated.
However, as rather well-defined but non-single magnetic domains 
(or domain states) form in complicated geometries,
it becomes difficult to characterize precisely magnetization reversal
involving each domain at small-length scales with either conventional 
magnetization loop measurements, such as magneto-optical Kerr effect 
(MOKE) magnetometry, or magnetic microscopy, such as magnetic force
microscopy (MFM).
Moreover, in large-area arrays typically covering areas of 
a few square millimeters, extracting overall domain structures during
reversal from microscopic images is clearly unreliable. 
Though diffracted MOKE measurement has been proposed recently
to deal with this problem, it is found to provide little 
quantitative information on magnetization reversal involving domain
formation and is limited to micrometer-length scales.\cite{dmoke_prb_02}

Such quantitative information is available using the technique of 
soft x-ray resonant magnetic scattering (SXRMS).\cite{kao_prl_90}
This technique exploits strong enhancement of the magnetic sensitivity 
of scattering intensities when incident circularly polarized soft
x-rays are tuned to an absorption edge of constituent magnetic atoms.
SXRMS has been used to study the magnetic structure in magnetic 
thin films\cite{sxrms_film} or periodic arrays of 
stripe domains or nanolines\cite{sxrms_line}. 
In this paper we present a simple scheme to extract quantitatively 
domain-specific magnetization reversal for nanomagnet arrays from 
SXRMS measurements.
For this purpose, sample rocking curves, yielding in-plane diffraction scans,
have been measured and analyzed on the basis of our previous work performed in 
the hard x-rays.\cite{lee_dot}
In order to obtain magnetic information, 
SXRMS peak intensities were measured by varying the applied field  
at different diffraction orders, whose scattering structure factors are 
different. 
This allows us to determine directly the magnetization reversal of each magnetic
domain using a simple linear algebra. 
The basic idea of incorporating such nonuniform magnetic domains 
in scattering theory has been explored in our previous work on
polarized neutron scattering.\cite{lee_neutron}    
       
For this study an array of permalloy (Ni$_{80}$Fe$_{20}$) square rings 
was fabricated by a combination of e-beam lithography and lift-off techniques.  
A standard silicon wafer was spin-coated with a double-layer 
positive-type e-beam resist, and  
the resist layer was then patterned by e-beam lithography.
A 20-nm-thick permalloy film was deposited  
onto it using an electron-beam evaporator in a vacuum of about $10^{-8}$ Torr.  
The as-deposited unpatterned film was magnetically soft with coercive and 
uniaxial anisotropy fields of a few Oersteds.  
Finally, after ultrasonic-assisted lift-off, 
the square rings were arranged in an array of 2 $\times$ 2 mm$^2$.

Experiments were performed at sector 4 of the Advanced Photon Source.\cite{sector4}
Polarized soft x-rays at the beamline 4-ID-C were generated by a novel
circularly polarized undulator that provided left- and
right-circular polarization switchable on demand at a polarization $> 96$\%.
The photon energy was tuned to the Ni L$_3$ absorption edge (853 eV) to 
enhance the magnetic sensitivity.
While a vacuum compatible sample stage was rotated,  
the diffracted soft x-ray intensities were collected by
a Si photodiode detector with a fixed angle of 12.8$^{\circ}$.
This sample rocking scan, where the incident and exit angles $\theta_i$ and
$\theta_f$ were varied with the total scattering angle $(\theta_i + \theta_f)$
fixed, yielded a transverse $q_x$ scan at a fixed $q_z$ value (see Fig. 1).
The angular resolution was defined by a pinhole between the sample and
detector to be about 0.03$^{\circ}$.   
The sample was mounted in the gap of an eletromagnet that provides fields
in the scattering plane of up to $\pm 800$ Oe.
\begin{figure}
\epsfxsize=7.5cm
\centerline{\epsffile{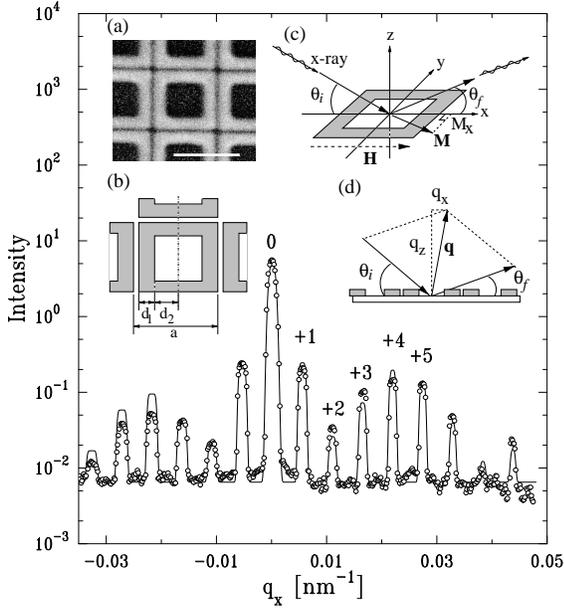}}
\caption{Diffraction intensities of the rocking scan along the $q_x$ direction
at $q_z = 0.955$ ${\rm nm}^{-1}$ from a square ring array. 
Circularly polarized soft x-rays were used and the photon energy was tuned to
the Ni L$_3$ absorption edge (853 eV).
Circles represent measurements, and lines represent the calculations.
Insets: (a) scanning electron micrograph and (b) schematic of the square ring array studied. 
The scale bar is 1 $\mu$m.  
(c) and (d) show schematics of the scattering geometry,
where $\theta_i$ and $\theta_f$ are the incidence and exit angles, and
${\bf H}$ and ${\bf M}$ are the applied field and magnetization, respectively.} 
\end{figure}  

Figure 1 shows the diffraction intensities of the sample rocking scan 
measured as a function of $q_x$ at $q_z = 0.955$ nm$^{-1}$ from the square ring array
with the saturation field.
Diffracted intensities show peaks corresponding to an array period of 1.151 $\mu$m.
Following Refs. \onlinecite{lee_dot,lee_neutron}, the diffracted intensity $I$
can be expressed in the kinematical approximation as
\bb
I(q_x ; H) &=& |F(q_z)|^2 \sum_{n_x, |n_y|} \Bigl|
       \rho_C^{} F_C \bigl(n_x,|n_y|\bigr) \nonumber \\
       &+& \rho_M^{}  F_M \bigl( n_x, |n_y|; H\bigr) \Bigr|^2
       {\cal R}\bigl(q_x; n_x, |n_y|\bigr),
\ee            
where $H$ is the applied field, $F(q_z)$ is the 1D form factor along $q_z$ direction 
and consequently a constant value for a fixed $q_z$, and 
$F_{C(M)}$ and $\rho_{C(M)}$ are the charge (magnetic) form factors on the $q_x-q_y$ plane
and the charge (magnetic) contributions to 
the total atomic scattering amplitude, respectively. 
Near resonance energies $\rho_M$ is proportional to the vector product 
$({\bf\hat{e}}^{\ast}_f\times{\bf\hat{e}}_i)\cdot{\bf M}$,\cite{kao_prl_90}
where ${\bf\hat{e}}_i$ and ${\bf\hat{e}}_f$ are unit
photon polarization vectors for incident and scattered waves, respectively, 
and ${\bf M}$ is the magnetization vector. 
For circularly polarized beams, this vector product reduces approximately 
to the component $M_x$ in the inset (c) of Fig. 1.   
Therefore, the magnetization referred to hereafter represents
strictly the parallel component to the $x$-axis or the field direction 
($H\parallel \hat{x}$ in this study) of the magnetization vector. 
$n_x$, $n_y$ are indices for Bragg points in the reciprocal 
space with the relationship of $q_{x,y} = (2\pi/a)n_{x,y}$,  
where $a$ is the period of the array.  
Since the resolution function ${\cal R}$ is a long thin ellipse 
oriented in the $q_y$ direction, nonzero $n_y$ values  
should also be taken into account for a $q_x$-scan performed 
at $q_y = 0$.\cite{lee_neutron,lee_dot} 
We also note that the evolution of the peak widths of the different
diffracted orders as a function of $q_x$, as shown in Fig. 1, was 
calculated using the model proposed by Gibaud $et$ $al.$\cite{gibaud_jphy_96}

For a square ring, the charge form factor $F_C(n_x, |n_y|)$ can be expressed by
\bb
F_C \bigl( n_x, |n_y|\bigr) 
&=& {\cal C} \biggl[ {\rm sinc}\Bigl( n_x (\gamma_1 + \gamma_2)\Bigr)
		   {\rm sinc}\Bigl( |n_y| (\gamma_1 + \gamma_2)\Bigr)
	\nonumber \\
	&-& \frac{\gamma_2^2}{(\gamma_1+\gamma_2)^2}
		   {\rm sinc}(n_x \gamma_2) 
		{\rm sinc}\Bigl(|n_y|\gamma_2\Bigr) \biggr],
\ee
where ${\cal C}$ is 1 for $|n_y|=0$ and 2 for $|n_y|\neq 0$,
and ${\rm sinc}(x)= \sin(x)/x$.    
$\gamma_{1,2} = 2\pi d_{1,2} /a$, where $d_1$ and $d_2$ are the width of the ring and
half of the inner square size, respectively, as shown in inset (b) of Fig. 1.   
Since $F_C$ and $F_M$ are the same functional forms 
for a saturated uniform magnetization, the diffracted intensities can be 
calculated using Eqs. (1) and (2) and are shown as the solid line in Fig. 1.
From the best fit, $d_1$ and $d_2$ were estimated to be $162\pm 4$ and $377\pm 4$
nm, respectively, and subsequently the gap between rings was 73 nm.
\begin{figure}
\epsfxsize=7cm
\centerline{\epsffile{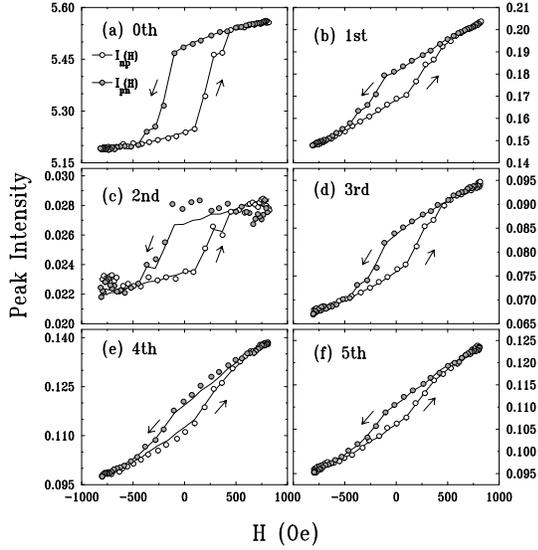}}
\caption{SXRMS magnetic hysteresis loops (circles) measured 
at several diffraction peaks, indicated by numbers in Fig. 1.  
The solid lines represent the calculated hysteresis loops.} 
\end{figure}  

Figure 2 shows SXMRS peak intensities measured at various
diffraction orders while field cycling.  
All field dependencies show magnetic hysteresis loops
but with different features.
This is due to different magnetic form factors for different
diffraction orders, which reflect nonuniform domain formation
during magnetization reversal, as pointed out 
in diffracted MOKE studies.\cite{dmoke_prb_02}
Therefore, the magnetic form factor $F_M$ in Eq. (1) should be expressed by
the sum of the contributions of all possible magnetic domains, i.e.,
\bb
F_M \bigl( n_x, |n_y|; H\bigr) 
= \sum_l m_l (H) F^{(l)}_M (n_x, |n_y|),
\ee
where $m_l(H)$ represents the field-dependent magnetization of the $l$-th domain,
which is the quantity of interest.    
It is noticeable that the magnetic factor in Eq. (3) can be factorized into
the field- and structure-dependent factors.
However, the domain-specific magnetization $m_l(H)$ cannot be directly extracted
from measured SXRMS hysteresis loops because, as described in Eq. (1), 
the diffracted intensities are the absolute square of the sum of the structural
and magnetic contributions.  
\begin{figure}
\epsfxsize=7cm
\centerline{\epsffile{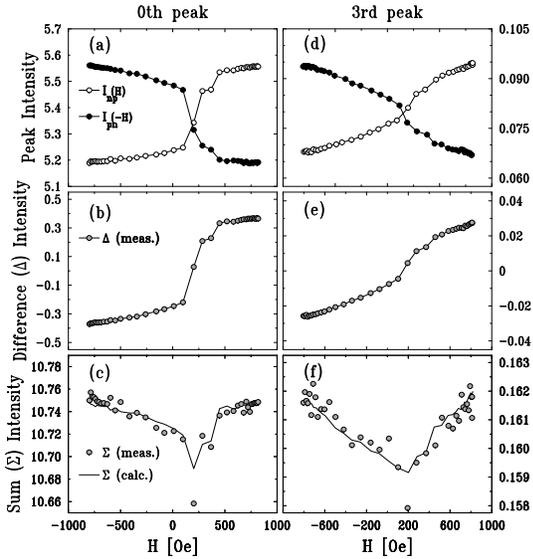}}
\caption{(a) and (d) SXRMS hysteresis loops with positive-to-negative intensities ($I_{pn}$)
flipped with respect to $H=0$ for zeroth- and third-order peaks, respectively.
Difference [(b) and (e)] and sum [(c) and (f)] intensities between 
$I_{np}(H)$ (negative-to-positive) and flipped $I_{pn}(-H)$ intensities 
as a function of the applied field.  
The solid lines in (c) and (f) represent the calculations.}
\end{figure}  

To tackle this problem, we considered the difference between the field-dependent 
intensities of $I_{np}(H)$ and $I_{pn}(-H)$, where $I_{np}$, $I_{pn}$ 
represent the intensities measured
while the field is swept along the negative-to-positive and positive-to-negative
directions, respectively, and 
$I_{pn}(-H)$ represents the intensities flipped from 
$I_{pn}(H)$ with respect to $H=0$, as shown in Fig. 3. 
Assuming that the magnetization reversal has inversion symmetry about the origin, 
only the difference between $I_{np}(H)$ and $I_{pn}(-H)$ is 
the opposite sign of $m_l(H)$ in Eq. (3).
However, this does not mean that they are symmetrical to a certain horizontal
line in Fig. 3 (a) or (d) because the intensities contain quadratic terms
to the magnetization $m_l(H)$, as described in Eq. (1), consequently
leading to a nonsymmetry to the origin (or the center of mass) 
of SXRMS hysteresis loops in Fig. 2.
This effect is clearly seen in the sum intensities of Figs. 3 (c) and (f) and
will be an important characteristic of scattering-based hysteresis loops. 
However, in turn, these quadratic terms can be 
ruled out by taking the difference intensities, which are, as a result, 
linearly proportional to the magnetization.

These difference intensities at the $n_x$-order peak can be 
expressed from Eqs. (1) and (3) as
\bb
&&\Delta_{n_x} (H) \equiv I_{np}(n_x; H) - I_{pn}(n_x; -H) \\
&&~~~~~~~~~~~= 4 {\rm Re} [\rho_C^{} \rho_M^{\ast}] {\cal R}_{n_x} |F(q_z)|^2 \nonumber \\
&&~~~~\times \sum_l m_l (H) 
\sum_{|n_y|=0}^{\infty} F_C \bigl(n_x,|n_y|\bigr)
F_M^{(l)}\bigl(n_x,|n_y|\bigr) {\cal R}_{n_y}, \nonumber
\ee
where ${\cal R}_{n_x}$ and ${\cal R}_{n_y}$ represent resolution functions evolving $n_x$ and
$n_y$ indices, respectively, into which ${\cal R}(n_x,|n_y|)$ in Eq. (1) 
can be factorized.
If we further normalize $\Delta_{n_x}$ by its maximum intensity with 
a saturation magnetization $m_s$, 
we can obtain a set of linear equations as
\bb
\frac{\Delta_{n_x}(H)}{|\Delta_{n_x}^{\rm max}|} 
= \sum_l B_{n_x l} \frac{m_l(H)}{m_s}, 
\ee
where
\bb
B_{n_x l} 
= \frac{\sum_{|n_y|} F_C \bigl(n_x,|n_y|\bigr) 
		F_M^{(l)}\bigl(n_x,|n_y|\bigr) {\cal R}_{n_y}}
	    {\sum_{|n_y|} F_C^2 \bigl(n_x,|n_y|\bigr) {\cal R}_{n_y}}. 
\ee
Here we used the relationship of 
$\sum_l^{all} F_M^{(l)}\bigl(n_x,|n_y|\bigr) = F_C \bigl(n_x,|n_y|\bigr)$.
Applying linear algebra, the normalized magnetizations 
$m_{l=1,\cdot\cdot\cdot,N}(H)/m_s$ 
of $N$ domains can be finally obtained directly from the normalized 
difference intensities $\Delta_{n_x}(H)/|\Delta_{n_x}^{\rm max}|$ measured
at $N$ different $n_x$ orders by taking the inverse of $N\times N$ matrix $B_{n_x l}$.  

In principle, the field-dependent intensities measured at 
semi-infinite numbers of orders can thus be used to determine the magnetization
reversal of each infinitesimal cell in a unit nanomagnet.
However, this is practically restricted due to finite measurable diffraction 
peaks and a high symmetry of experimental geometry.
The latter gives rise to a strong dependence between $F_M^{(l)}\bigl(n_x,|n_y|\bigr)$
with different domain $l$ or diffraction order $n_x$, and, as a result,
makes the matrix $B_{n_x l}$ singular and noninvertible.
To lower the geometrical symmetry of experimental configuration, 
we can use a vector magnetometry setup by rotating
the sample/electromagnet assembly with respect to 
the incident photon direction.\cite{lee_hole}
Since SXRMS intensities are proportional to the component of the magnetization vector
along the projected incident photon direction onto the sample surface,     
this approach can also provide vectorial information about domain-specific magnetization.

In our setup, where both incident beam and field directions are parallel to 
one of the sides of the square rings, there may be four characteristic domains,
as shown in the insets of Fig. 4.
Each domain consists of two or four subdomains, whose structure-dependent 
form factors in Eq. (3) are identical, and, therefore, its magnetization represents
an average value over subdomains.
The explicit expressions of the structure-dependent magnetic form factors $F_M^{(l)}$ 
of these four domains for $(n_x, |n_y|)$ diffraction order in Eq. (3)
will be presented elsewhere.\cite{lee_elsewhere}
We note that these domains have been chosen to minimize sigularity of 
the matrix $B_{n_x l}$
and may not be thus energetically viable. 
Nevertheless, this scheme can unambiguously provide information about domain
formation by considering snapshots of the resultant domain-specific magnetizations at
each field.   
\begin{figure}
\epsfxsize=7.5cm
\centerline{\epsffile{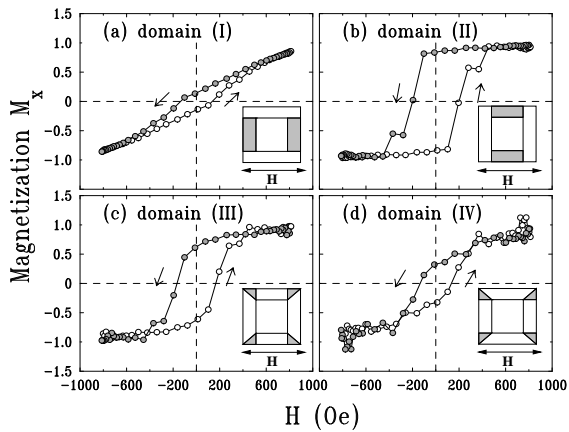}}
\caption{Extracted magnetization reversals along the field direction 
of four types of characteristic domains, which are depicted as gray-filled regions 
in the insets, from SXRMS hysteresis loops at zeroth-,
first-, third-, and fifth-order peaks.  } 
\end{figure}  

To construct a $4\times 4$ matrix $B_{n_x l}$ for four magnetic domains,
at least four different peak intensities are required, and fifteen combinations
can be allowed for six measured peaks, as shown in Fig. 2.
However, we excluded the second peak due to its bad statistics and also 
some other combinations due to relatively small values of the determinants of
their matrices $B_{n_x l}$, leading to a singularity of the matrix.
For the optimum combination for non-singularity, zeroth-, first-, third-,
and fifth-order peaks were then chosen. 
Figure 4 shows the finally extracted magnetization reversals 
for each domain using Eqs. (5) and (6).  
All magnetizations in Fig. 4 are normalized by the saturation and 
represent the components projected along the field direction, as discussed
above, of the magnetization vectors.
To confirm these results, the sum intensities in Figs. 3(c) and 3(f) 
and the SXMRS hysteresis loops for all observed peaks in Fig. 2 
were also generated by substituting the results in Fig. 4 into Eq. (1). 
These calculations (solid lines) show a good agreement with the measurements.

A remarkable feature in the extracted magnetization reversals is
that while the domain (I) perpendicular to the field rotates coherently,
the parallel domain (II) is strongly pinned.
Interestingly, this is similar to the domain behaviors 
in the antidot arrays,\cite{hole_apl_97,lee_hole} 
whose geometry resembles the square
ring array except for narrow gaps between rings.
On the other hand, the domain (II) clearly shows plateaus, which 
have been observed generally in circular or octagonal ring magnets and 
are attributed to the vortex state.\cite{ring_prl_01}
A detailed discussion is beyond the scope of this paper and will be presented
elsewhere.\cite{lee_elsewhere} 
 
In summary, we successfully demonstrated that domain-specific magnetization
reversals can be extracted directly from SXRMS 
hysteresis loops measured at various diffraction orders.        
Extracted domain-specific magnetization reversals are expected to provide
a new insight into magnetic switching mechanism on nanofabricated arrays.
Future studies, exploiting the element-selectivity and vector magnetometry
setup, will provide further three-dimensional information 
in nanofabricated multilayers such as giant magnetoresistance and 
pseudospin valve structures.  

Work at Argonne is supported by the U.S. DOE, Office of Science, 
under Contract No. W-31-109-ENG-38.		
V.M. is supported by the U.S. NSF, Grant No. ECS-0202780.


\begin{references}   
\bibitem{record_jpd_02} A. Moser {\it et~al.},
	J. Phys. D: Appl. Phys. {\bf 35} R157 (2002).
\bibitem{spintronics_ieee_03} S. Parkin {\it et~al.},
	P. IEEE {\bf 91}, 661 (2003).
\bibitem{disk_jpd_00} R. P. Cowburn,
	J. Phys. D: Appl. Phys. {\bf 33} R1 (2000).
\bibitem{ring_prl_01} J. Rothman {\it et~al.},
	Phys. Rev. Lett. {\bf 86}, 1098 (2001);
 	S. P. Li {\it et~al.}, 
	{\it ibid.} {\bf 86}, 1102 (2001).
\bibitem{hole_apl_97} R. P. Cowburn, A. O. Adeyeye, and J. A. C. Bland,
	Appl. Phys. Lett. {\bf 70}, 2309 (1997).
\bibitem{dmoke_prb_02} I. Guedes {\it et~al.},
	Phys. Rev. B{\bf 66}, 014434 (2002).
\bibitem{kao_prl_90} C. Kao {\it et~al.},
	Phys. Rev. Lett. {\bf 65}, 373 (1990).
\bibitem{sxrms_film} J. M. Tonnerre {\it et~al.},
	Phys. Rev. Lett. {\bf 75}, 740 (1995);
	J. F. MacKay {\it et~al.}, 
	{\it ibid.} {\bf 77}, 3925 (1996);
	Y. U. Idzerda, V. Chakarian, and J. W. Freeland,
	{\it ibid.} {\bf 82}, 1562 (1999);
	J. W. Freeland {\it et~al.},
	Phys. Rev. B{\bf 60}, R9923 (1999);
	J. B. Kortright {\it et~al.},
	{\it ibid.} {\bf 64}, 092401 (2001).
\bibitem{sxrms_line} H. A. D\"{u}rr {\it et~al.},
	Science {\bf 284}, 2166 (1999);
	K. Chesnel {\it et~al.}, 
	Phys. Rev. B{\bf 66}, 024435 (2002).
\bibitem{lee_dot} D. R. Lee {\it et~al.},
	Appl. Phys. Lett. {\bf 82}, 982 (2003).
\bibitem{lee_neutron} D. R. Lee {\it et~al.},
	Appl. Phys. Lett. {\bf 82}, 82 (2003).
\bibitem{sector4} J. W. Freeland {\it et~al.},  
	Rev. Sci. Instrum. {\bf 73}, 1408 (2001).
\bibitem{gibaud_jphy_96} A. Gibaud {\it et~al.},
	J. Phys. I {\bf 6}, 1085 (1996).
\bibitem{lee_hole} D. R. Lee {\it et~al.},
	Appl. Phys. Lett. {\bf 81}, 4997 (2002).
\bibitem{lee_elsewhere} D. R. Lee {\it et~al.}, in preparation.
\end{references}
\end{document}